\documentclass[10pt,aps,pra,reprint,showpacs,superscriptaddress,floats,floatfix,amsmath,amssymb]{revtex4-1}
\usepackage[pdfpagemode=None]{hyperref}
\usepackage{graphicx}
\usepackage{physics}
\usepackage{bbold,bm}
\usepackage{color}
\begin{document}
\title{{How to observe and quantify quantum-discorded states via correlations}}

\author{{Matthew A. Hunt}}
 \affiliation{School of Physics \& Astronomy, University of Birmingham, B15 2TT, UK}
 \author{Igor V. Lerner}
\affiliation{School of Physics \& Astronomy, University of Birmingham, B15 2TT, UK}
  \author{Igor V. Yurkevich}
\affiliation{School of Engineering \& Applied Science, Aston University, Birmingham B4 7ET, UK}
 \author{Yuval Gefen}
 \affiliation{Department of Condensed Matter Physics, The Weizmann Institute of Science, Rehovot 76100, Israel}


\begin{abstract}
 Quantum correlations between parts of a composite system most clearly reveal  themselves through entanglement. Designing, maintaining, and controlling  entangled systems is very demanding, which raises the stakes for understanding the efficacy of entanglement-free,  yet quantum, correlations,  exemplified by \emph{quantum discord}.
Discord is defined via conditional mutual entropies of parts of a composite system,
and  its direct measurement  is hardly possible even via full tomographic characterization of the system state.  Here we design a simple protocol to detect and quantify quantum discord in an \emph{unentangled} bipartite system.
Our protocol relies on a characteristic of discord that can be extracted from \emph{repeated} direct measurements of certain correlations between subsystems of the bipartite system. The proposed protocol opens a way of extending experimental studies of discord  to electronic systems but can also be implemented in quantum-optical systems.
\end{abstract}

\maketitle
\section{Introduction}

While quantumness of correlations between the parts of a system in a pure state is fully characterized by their entanglement (see Ref.~\onlinecite{Horodecki4} for reviews), mixed states may possess quantum correlations even if they are not entangled. The quantumness of the correlations is properly described in terms of
quantum discord \cite{Zurek:01,Vedral:01,discord-n1}   which is a discrepancy between  quantum versions of two classically equivalent expressions for mutual entropy in bipartite  systems (see Ref.~\cite{Modi:12,DiscordRev-18b,DiscordRev-18} for reviews).
Any entangled  state of a bipartite system is discorded, but  discorded states may be non-entangled. Although it is entanglement which is usually assumed to be
the key resource for quantum information processes, it was suggested that quantum enhancement of the efficiency of data processing can be achieved in deterministic quantum computation with one pure qubit which uses mixed \emph{separable} (i.e.\ non-entangled) states \cite{Knill:98,*Knill-Nat:02,*Datta:05,*Cable:16}. In such a process, which has been experimentally implemented \cite{White:08}, the nonclassical correlations captured by   quantum discord are responsible for   computational speedup \cite{Datta:08}. Quantum discord was also shown to be the necessary resource for   remote state preparation  \cite{Dakic-NatPhys}, and for the distribution of quantum information to many parties  \cite{Zurek:13,*Horodecki:15}. Unlike entanglement, discord is rather robust against decoherence \cite{Mazzola:10}.
Thus, along with   entanglement,   quantum discord can be harnessed for certain types of quantum information processing.

Despite increasing evidence for the relevance of quantum discord, \emph{quantifying} it  in a given quantum state is a challenge. Even full quantum state tomography would not suffice, since determining discord   requires minimizing a conditional mutual entropy over a full set of projective measurements.  Moreover, even computing discord is very difficult (it has been proven to be NP-complete \cite{Huang_2014}). An alternative, geometric measure of  discord \cite{Vedral:10,GDDiscordWitn,DiscordWitness2,Brodutch:2012}  has been successfully implemented experimentally \cite{DiscordWitness1,*Silva:13,*Benedetti:13}. However, geometric discord also faces serious problems. For example,  it can increase, in contrast to the original quantum discord,  even under trivial local
reversible operations on the passive part of the bipartite system \cite{Piani:12} (note, though, the proposal of Ref.~\cite{PhysRevA.88.012120} to mend this deficiency). Most seriously, being a  non-linear function of the density matrix $\rho$, geometric discord can only be quantified via (full or partial) reconstruction of  $\rho$ itself. This severely limits its susceptibility to experiment in the many-body context.

In this paper we   propose a novel discord quantifier which would  overcome these fundamental difficulties and render quantum discord to be experiment-friendly  for many-body electronic systems, where it has not yet been observed. We present a
  protocol  to detect and characterize quantum discord of any unknown mixed state of a generic \emph{non-entangled} bipartite system, implemented in either electronic or photonic setup.  The protocol is based on direct repeated measurements of certain two-point correlation functions (which are linear in $\rho$ as any direct quantum-mechanical observable).  While discord cannot be detected by a single linear measurement \cite{Rahimi:10,DiscordRev-18}, we show  how \emph{repeated} measurements   allow one to both detect a discorded state and build its reliable quantifier.

 {In the next section we describe the principle steps of the proposed protocol. In Section \eqref{section3}, we demonstrate how to implement our protocol in an electronic bipartite system built on integer quantum Hall devices and prove that it provides a reliable discord witness.  In Section \eqref{section4}, we illustrate how the protocol works by applying it to a few specified states and propose a new discord quantifier based in this protocol.  Finally, in Section \eqref{section5}, we explain how the protocol should be applied to an unknown state.}

\section{Principal steps of the protocol\label{section2} }

Here we describe how to detect quantumness in unentangled states of a bipartite system via correlations.
 A generic non-entangled bipartite system in a mixed state
 is described by the   density matrix \cite{Werner:89}
\begin{align}\label{sepstate}
 \rho^{AB}&=\sum_{\nu=1}^Mw_\nu\rho^A_\nu\otimes\rho^B_\nu,
 \end{align}
where  the classical probabilities $w_\nu$ add up to $1$, and each $\rho_\nu^{A,B} $ describes a pure state of the appropriate subsystem, so that they can be parameterized as $\rho_\nu^{A}=  \ket{{A}_\nu}\bra{A_\nu}  $ (and similarly for $B$).
 It turns out \cite{Vedral:10,Ferraro:10,Modi:12}   that  the mixed state (\ref{sepstate}) is $A$-discorded \footnote{Quantum discord is not necessarily symmetric: one can record discord in one (active) subsystem ($A$) of a bipartite system while the other (passive) subsystem ($B$) might be either discorded or not.} independently of  $\rho_\nu^B$, \emph{unless} the set $\{ \ket {A _\nu} \} $ forms an orthogonal basis. In order to detect and quantify $A$-discord, we propose to utilize this property of state (\ref{sepstate}). {Let us describe principal steps of the proposed protocol.}

 \begin{enumerate}
   \item {Prepare a bipartite system in the mixed input state described by matrix \eqref{sepstate}.
 }
   \item {Let the system  evolve  into an out-state described by density matrix \\[3pt]
$\phantom{a} \qquad \qquad\qquad \widetilde{\rho}^{AB}={\mathsf{S\rho^{AB} S}}^\dagger  $  \\[3pt] with the unitary evolution matrix ${\mathsf{S=S^{{\mathrm{A}}}\otimes S^{{{\mathrm{B}}}}  }}$. }
   \item {Test a post-evolution rotation of the $A$-basis by allowing subsystem $A$ to evolve further through a detecting contour so that
$ {\mathsf{S}}^A \to {\mathsf{S}}_d({\phi_d }){\mathsf{S}}^A $. For simplicity, we assume that the evolution through the detecting contour is characterized by a single phase factor, $\phi _d$.
\item Make correlated projective measurements, $\Pi_{A,B}$,
 on both subsystems (with the detecting contour included in $A$):\\[3pt]
$\phantom{a} \qquad\qquad \qquad K_{\phi_d}=\Tr \!\Big[ {\Pi}_A\,{\Pi}_B\,  \widetilde{{\rho}}^{AB}
 \Big]  $
 }
 \item {Repeat the measurements with changing $\phi_d$ to get the interference pattern\\[3pt]
$\phantom{a}\qquad\qquad\qquad K_{\phi_d}=\mathcal{C}+\left(\mathcal{A}e^{i\phi _d}+{\mathrm{c.c.}}   \right)$.
\item Extract the interference visibility,\\[3pt]
${\phantom{a}}{}\qquad\mathcal{V=\dfrac{|A|}{|C|}}=\dfrac{\max [K_{\phi_d}]-\min [K_{\phi_d}]}{\max [K_{\phi_d}]+\min [K_{\phi_d}]}$
 }
 \end{enumerate}
 {The visibility $\mathcal{V}$ is a function of parameters of the two subsystems encoded in $\mathcal{A}$ and $\mathcal{C}$. The thrust of the proposed protocol is in the fact that the lines of zero visibility for input state \eqref{sepstate} with no $A$-discord remain the same with changing  parameters of subsystem $B$.  Hence a dependence of zero-visibility lines on parameters of $B$ for some input state signifies  $A$-discord of this  state. }

 {%
 The proposed protocol can be applied to any bipartite system. However, to be concrete, we will focus at its particular implementation in a quantum-Hall based two-qubit interferometry setup (or, equivalently, in an optically based interference setup) which can be experimentally realized.  We will prove that discord is reliably witnessed by the dependence of zero-visibility lines on parameters of the passive subsystem. Then we employ this dependence of visibility to quantify discord.
 }

 {We will illustrate how the protocol works using as examples some known bipartite states. However, it is aimed at implementing to unknown states, and we will describe in detail how this can be done in experiment.}

\section{Protocol in   detail for Quantum-Hall based setup\label{section3} }
It is well known \cite{DiscordRev-18} that a separable state can be prepared by local operations and classical communications. Here we propose a particular way of preparing such a state in a solid state setup {and explain in detail how the protocol described in the previous section works in this setup.}

\begin{figure}
\begin{center}
\includegraphics[width=0.47\textwidth]{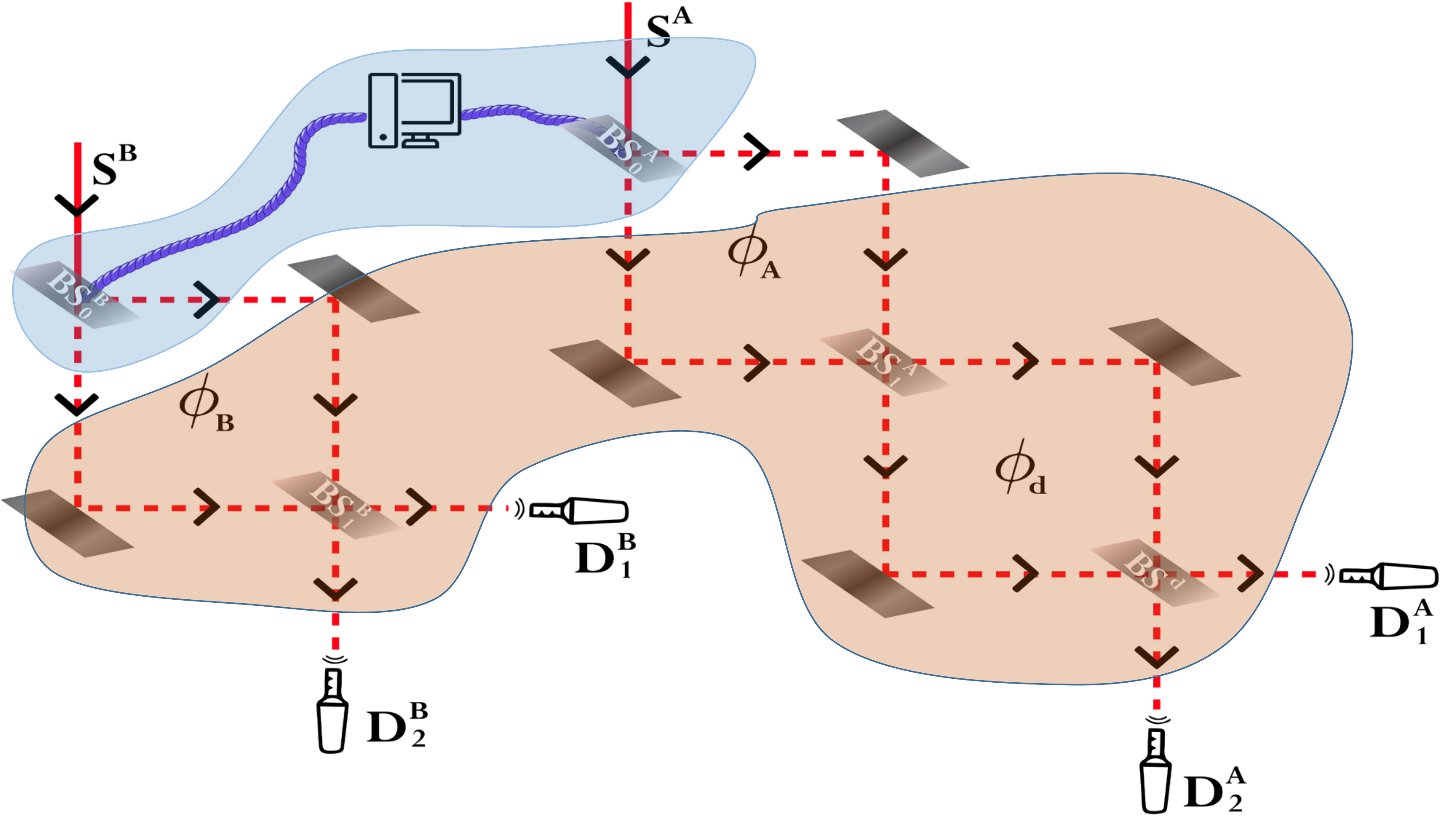}
\end{center}
\caption{\label{MZI} Proposed setup of the bipartite  system made of two Mach-Zehnder interferometers, {MZI$^A$ with the  phase difference $\phi _A$, and MZI$^B$ with  $\phi _B$}. The light blue (light brown) area represents the state-preparation (the state evolution and discord measurement) part of the protocol. Electrons from sources $S^A$ and $S^B$ enter  beam-splitters  BS$_0^{A} $ and BS$_0^B$,    whose random transparencies  are synchronized by a classical computer allowing the creation of mixed states of the form given by Eq. (1). The final state is controlled by transparencies of beam-splitters  BS$_1^{A} $ and BS$_1^B$ and phases   $\phi _{A,B}$, and  is recorded at any pair of detectors $D^A_i$ and $D^B_i$ (with $i=1$ or $2$).  Varying the phase difference $\phi _d$  in the third, detecting  MZI$^d$,  would allow one to identify a state with no $A$-discord as one for which the interference pattern  is suppressed for certain parameters of subsystem $A$ and remains suppressed for any tuning of subsystem $B$ (without adjusting $A$ any further), as illustrated below in Fig.~\eqref{fig2}.}
\end{figure}

A two-qubit bipartite system with a mixed state of Eq.~\eqref{sepstate}  can be implemented with the help of two Mach-Zehnder interferometers, MZI$^A$ and MZI$^B$, corresponding to subsystems $A$ and $B$ (cf.\ Fig.~\ref{MZI}). Such a system can be realized as an electron-based setup in a quantum Hall geometry, where the arms of the MZIs are constructed via a careful design of chiral edge modes, and quantum point contacts (QPC) act as effective beam-splitters (BS) \cite{Heiblum:03,Weisz1363,Heiblum:15}. It can also be realized as a photonic device using standard interferometry.

Each interferometer is in a quantum superposition of up, $\ket{\uparrow}$,  and down, $\ket{\downarrow}$, states corresponding to a particle  transmitted through the upper or lower arm of the appropriate MZI. {Such a superposition in subsystem $A$ can be parameterized as
 \begin{align}\label{X-nu}
    \ket{A_\nu}\equiv \ket{\theta_\nu,\,\phi_\nu } =\cos \tfrac{1}{2}\theta_{ \nu} \ket{\uparrow}+ {\mathrm{e}}^{i\phi _\nu } \sin\tfrac{1}{2} \theta_{ \nu}\ket{\downarrow},
\end{align}
with $\ket{0,0}\equiv \ket{\uparrow}  $, $\ket{\pi ,0}\equiv \ket{\downarrow}  $, and $\ket{\pm\frac{\pi }{2},0}\equiv \ket{\pm}=[{\ket{\uparrow}\pm\ket{\downarrow}  }]/\sqrt{2}.  $
Notations for subsystem $B$  are similar (we suppressed indices $A,B$ for now).
The coefficients in each superposition are  determined  by the gate-controlled transparency/reflection of the appropriate BS (with the corresponding  amplitudes parameterized as $t =\cos \frac{1 }{2}\theta_\nu$ and $r ={\mathrm{e}}^{i\phi _\nu} \sin\frac{1  }{2}\theta_\nu$),} where index $\nu$ labels the states defining density matrices $\rho_\nu^{A,B} $ in Eq.~\eqref{sepstate}.
Such a mixed state  can be  created  with the help of a \emph{classical} computer  that simultaneously and randomly switches  transparency/reflection   of  BS$^A_0$ and BS$^B_0$  between $n$ values.  The probabilities  $w_ \nu$  in this equation are now  proportional to the time of the pair of BS$_0$ having the appropriate transparencies, provided that the output on the detectors $D^B_{1,2} $ and $D^A_{1,2} $ is averaged over  time intervals much longer than the switching time. {This implements the first step of the protocol described above.}

{Step 2 of the protocol is a unitary evolution of the prepared mixed state  through the system, $\widetilde{\rho}^{AB}={\mathsf{S\rho^{AB} S}}^\dagger  $. Here each of the scattering matrices ${\mathsf{S}}^{A,B} $ through MZI$^{A,B} $ in ${\mathsf{S=S^{\mathrm{A}} \otimes S^{\mathrm{B}} }}$ includes first a phase difference $\phi _{A,B}$ between the $\ket{\uparrow} $ and $\ket{\downarrow} $  arms of the appropriate MZI, which is controlled by the  Aharonov--Bohm flux (measured in units of the quantum flux, $hc/e$), and scattering through the second set of beam splitters, BS$^B_1$ and BS$^A_1$. We can  parameterize these scattering matrices as the product of that corresponding to the beam-splitter and the phase difference accumulated on the opposite arms,}
\begin{align}\label{SB}
 \mathsf{S}^A&=\left(\begin{matrix}
	                        r_A & t_A \\
                            -t^*_A & r _A^*\\
                            \end{matrix}\right)\, e^{\frac{i}{2} \sigma_3 \phi _A} ,
 \end{align}
  {and likewise for ${\mathsf{S}}_B$. The transmission (and thus reflection) amplitudes could be represented similarly to those in the input MZIs  as $t_A=\cos\frac{1 }{2}\alpha $ and $t_B=\cos \frac{1}{2}\beta $ (with the phase factors in $r_{A,B}$ absorbed by the Aharonov -- Bohm phase). In repeating measurements with the same input state of Eqs.~\eqref{sepstate} and \eqref{X-nu}, one can accumulate statistics by varying parameters of the scattering matrices.}

{Step 3 of the protocol is to  test, as described below, whether the basis $\{{\ket{A_\nu}}\} $ of the $A$-part of input state \eqref{sepstate} was orthogonal or not, i.e.\ whether the system does not or does have $A$-discord. To this end we allow the active subsystem $A$ to further evolve  through  third, \emph{detecting} MZI$^d$ attached to it (see \ Fig.~\ref{MZI}), so that the full unitary  $S$-matrix} that describes  independent evolution of the mixed in-state Eq.~\eqref{sepstate} through subsystems $A$,  $B$ can be represented as
\begin{subequations} {\label{SSd}\begin{align}\label{S}
    {\mathsf{S}}= {\mathsf{S}}^B\otimes\left({\mathsf{S}}^d\, {\mathsf{S}}^A\right).
 \end{align}
Matrix ${\mathsf{S}}^d$ has the same structure as ${\mathsf{S}}^A$, Eq.~\eqref{SB}. However, it is sufficient for the testing to choose a 50:50 beam-splitter in MZI$^d$, so that
 \begin{align}\label{Sd}
 {\mathsf{S}}^d=\tfrac{1}{2}({\mathbb1+i\sigma_2}){\mathrm{e}}^{i\phi_d\sigma_3/2}
,\end{align}}\end{subequations}
 {with the Aharonov -- Bohm phase $\phi _d$ remaining the only tuneable parameter of the detecting MZI. }

{In step 4, we choose a cross-correlation function that describes a simultaneous detection of  particles injected into   $A$ and $B$  at the  detectors $D^A_1$ and $D^B_1$, so that the corresponding projector operators are  ${\Pi_{A,B} }=\ket{\uparrow}\bra{\uparrow}
 $  in the appropriate space.  Hence, with the output density matrix $\widetilde{\rho}^{AB}= {\mathsf{S}}\,{\rho}^{AB}
\,{\mathsf{S}}^{\dagger}$ and $S$-matrix defined by Eqs.~\eqref{SSd} we have
 \begin{align}
   K_{\phi_d}=\Tr \!\big[ {\Pi}_A{\Pi}_B\,\widetilde{\rho}^{AB} \big] &= \tfrac{1}{2}\Tr_A\! e^{\frac{i}{2}\sigma_3\,\phi_d}\, \widetilde{\rho}^{A|B}\, e^{-\frac{i}{2}\sigma_3\,\phi_d}\!,\notag\\  \label{K}\\[-6pt]
   \widetilde{\rho}^{A|B}&={\mathsf S}^A\,\rho^{A|B}\,\big({\mathsf S}^ A\big)^{\dagger}.\notag
 \end{align}
Here $\widetilde{\rho}^{A|B} $ is   the  conditioned output density matrix of the active subsystem $A$. The corresponding input density matrix  $\rho^{A|B}$ resulting from tracing over passive subsystem $B$
can be written as}
\begin{align}\label{rhoBA}
 \begin{aligned}
 \rho^{A|B}&=\frac{1}{W_B}
 \sum_{\nu =1}^nw^B_{\nu}\rho^A_\nu,\\  w^B_{\nu} &\equiv w_{\nu} \Tr_B \big[\Pi_B{{\mathsf{S}}
^B\rho^{B}_\nu \big( {\mathsf{S}}^B\big) ^\dagger
}\big],\quad{W_B}\equiv
 \sum_{\nu =1}^nw^B_{\nu}.
 \end{aligned}
\end{align}

 {Steps 5 and 6 of the proposed protocol are exactly as described in Section \eqref{section2}. Due to interference between the $\ket{\uparrow} $ and $\ket{\downarrow} $ states     in  MZI$^d$,   correlation function \eqref{K} oscillates with the phase difference $\phi _d$. By changing $\phi _d$ in repeated measurements of $K_{\phi _d}$ (Step 5) one accumulates statistics to get the interference visibility function $\mathcal{V}$ (Step 6).
 In the present setup, $K_{\phi_d}$ is an implicit function of parameters $\alpha $ and $\phi _A$, and $\beta $ and $\phi _B$ that define the evolution matrices ${\mathsf{S}}^A$ and ${\mathsf{S}}^B$, respectively.
The visibility vanishes when  $K_{\phi _d}$  becomes  $\phi _d$-independent.} This happens when $\widetilde{\rho}^{A|B}$  in Eq.~\eqref{K} is diagonal, i.e.\ ${\mathsf{S}}^A\to{\mathsf{S}}_0^A$, the diagonalizing matrix   for $\rho^{A|B}$. Such a diagonalization is always possible so that the zero-visibility lines exist for any input state.

\begin{figure*}
\qquad\includegraphics[height=.43\textwidth]{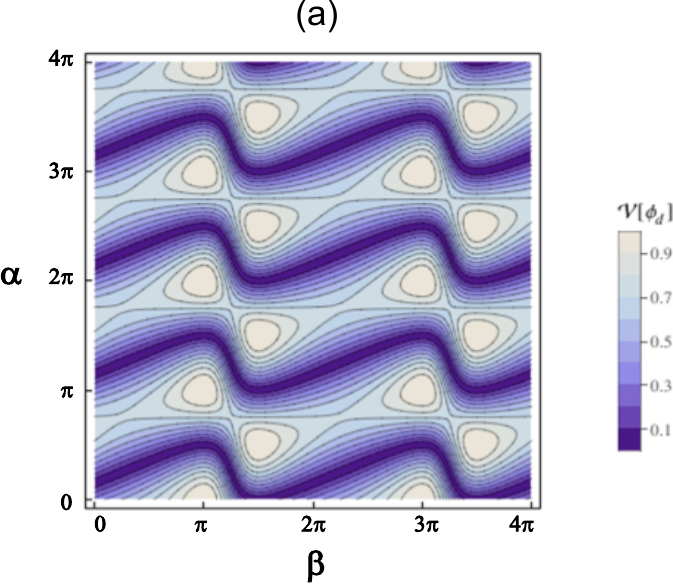}
\quad
\includegraphics[height=.43\textwidth]{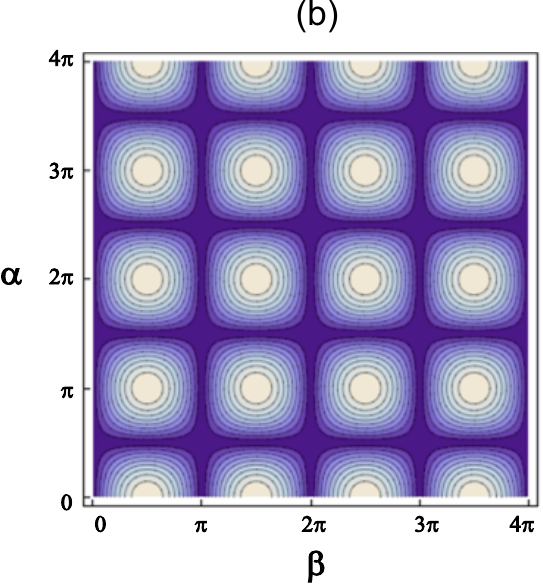}
\vspace*{-6pt}

\caption{\label{fig2}    A striking difference between (a)  discorded and (b)   non-discorded states: \emph{zero-visibility} (dark) lines are sensitive to changes in the state of passive subsystem $B$  (a) and  are independent of this changes in (b).     Here we use the symmetric in-states:   (a) $\rho^{AB} = \frac{1}{2}\left[ \ket{\uparrow\uparrow}\bra{\uparrow\uparrow}{\text+} \frac{1}{2} \ket{\bm {++}}\bra{\bm {++}}\right]   $  and  (b) $\rho^{AB} =\frac{1}{2} \left[\ket{\bm {++}}\bra{\bm {++}}{\text+}\frac{1}{2} \ket{\bm {--}}\bra{\bm {--}}\right]   $ with $\ket{\bm\pm}\equiv \frac{1}{\sqrt{2}}({\ket{\uparrow} \pm\ket{\downarrow} }) $.
Since the states $\ket{+} $ and $\ket{-} $ are  orthogonal whereas $\ket{+} $ and $\ket{\uparrow} $ are not,   these density matrices describe  a  discorded state (a) and a  non-discorded state (b), as explained after Eq.~\eqref{rhoBA}. Any continuous zero-visibility line in (b) can be chosen  for a quantitative characteristic of discord, Eq.~\eqref{W},   cf.\ Fig.~\ref{Discord}.}
\end{figure*}

{The central point of the proposed protocol is that such zero-visibility lines are independent of parameters of passive subsystem $B$ only if the $A$-discord vanishes. Now we prove this for the setup under consideration.}

{We begin with parameterizing input states $\ket {A_\nu} $ in subsystem $A$, Eq.~\eqref{X-nu}, via the unit vector $\bm a_\nu$ on the appropriate Bloch sphere}
\begin{align}
\label{n-nu}
  {\bm a}_{\nu}&=(\sin\theta_{\nu}\cos\phi_{\nu},\sin\theta_{\nu}\sin\phi_{\nu},\cos\theta_{\nu})\,,
 \end{align}
{ so that $\rho_\nu^A\equiv \ketbra {A_\nu}=\frac{1}{2}({\mathbb1+{\bm{a}_{\nu}\cdot{\bm{\sigma}}}})$ in the up-down basis where ${\mathbb{1}}=\ketbra\uparrow+\ketbra\downarrow$. Then we represent the conditioned density matrix $\rho^{A|B} $ in Eq.~\eqref{rhoBA} as}
\begin{align}\label{ABa}
    \rho^{A|B}&=\tfrac{1}{2}\Big[c\ketbra{{A}} -\big(c-W_B\big){\mathbb{1}}\Big], &{\bm{a}}&\equiv \frac{1}{c}\sum_{\nu} w_\nu^B{\bm{a}}_\nu,
\end{align}
{via  axillary unit vector \mbox{${\bm{a}} \equiv (\sin\vartheta \cos\varphi ,\sin\vartheta \sin\varphi ,\cos\vartheta )$} (with $c$ being the normalization constant), corresponding to the state $ \ket{A}=\cos \tfrac{1}{2}\vartheta  \ket{\uparrow}+ {\mathrm{e}}^{i\varphi  } \sin\tfrac{1}{2} \vartheta \ket{\downarrow}$.
From this representation follows that $\widetilde{\rho}^{A|B} $ in Eq.~\eqref{K} becomes diagonal when the diagonalizing matrix ${\mathsf{S}}^A_0$ obeys, up to a phase factor, the following equation   that defines zero-visibility lines:}
\begin{align}\label{S0A}
{\mathsf{S}}^A_0\ket{A}=\ket \uparrow\,\text{ or }\,\ket \downarrow
\end{align}

{Since the unitary matrix ${\mathsf{S}}^A$ rotates vectors on the Bloch sphere, the solutions to this equation correspond  to the rotations to the north, $\ket{\uparrow} $, or south, $\ket{\downarrow} $, pole are given by the angles $\phi _A=\varphi $ and $\alpha =\vartheta$, {or $\phi _A=-\varphi $ and $\alpha =\pi -\vartheta$}, in parameterization of Eq.~\eqref{SB}.}

{The angles $\vartheta$ and $\varphi $ are to be found from the definition of ${\bm{a}}$, Eq.~\eqref{ABa}. It follows from this definition that if the unit vectors ${\bm{a}}_\nu$ are either the same (so that $\rho^{AB}=\rho^A\otimes \rho^B  $)  or antiparallel (so that the appropriate $\ket{A_\nu} $ are orthogonal), then ${\bm{a}}$ does not depend on $w^B_\nu$, i.e.\ on the state of subsystem $B$ (parameterized by angles $\beta $  and $\phi _B$). It is straightforward to see the converse: if ${\bm{a}}$ is $B$-independent, then ${\bm{a}}_\nu$ are either the same or opposite vectors on the Bloch sphere, so that the corresponding states $\ket{A}_\nu $ either coincide (up to a phase factor), or orthogonal. But such groups of states are the only ones when the bipartite system of Eq.~\eqref{sepstate} has no $A$-discord.}

{Hence we have proved that the sensitivity of zero-visibility lines to a state of the passive subsystem is a reliable discord witness: $A$-discord is absent if and only if such sensitivity is absent. In the next session, after illustrating this with a few examples we demonstrate how to build a discord quantifier based on this sensitivity.}

\section{Correlation-based discord quantifier\label{section4}}

\begin{figure*}
\qquad\includegraphics[height=.43\textwidth]{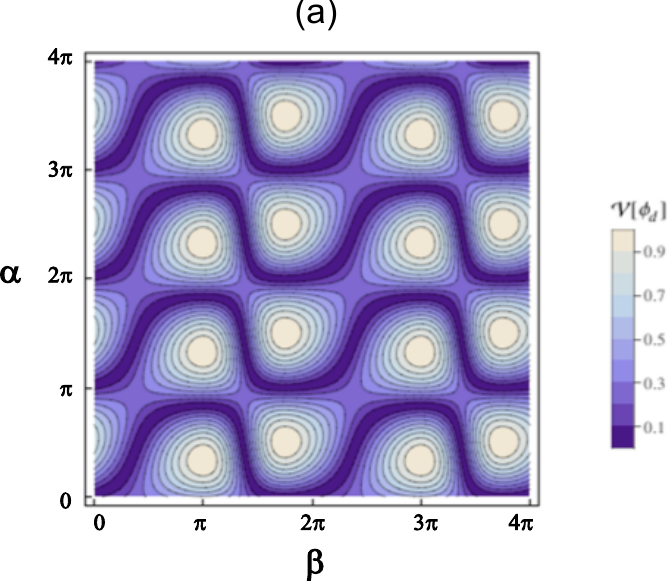}\quad
\includegraphics[height=.43\textwidth]{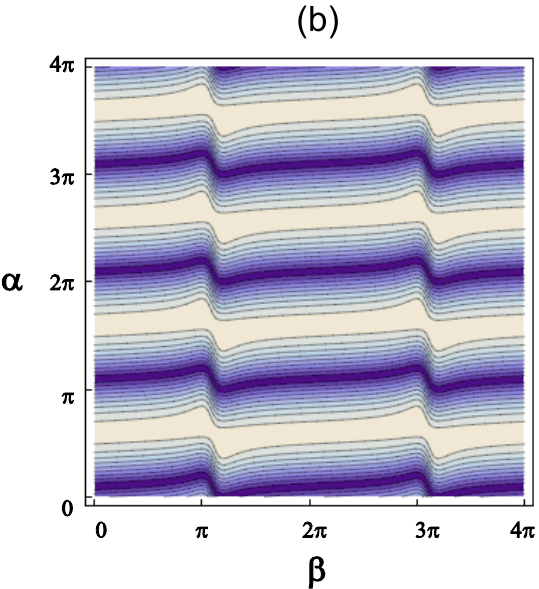}
\vspace*{-6pt}

\caption{ \label{fig3}  The visibility plots for $\rho^{AB}=\frac{1}{2}\ket{\uparrow\uparrow}\ket{\uparrow\uparrow} +\frac{1}{2}\ket{\theta\theta}\bra{\theta\theta}$ where in (a)  $\theta=\frac{5}{6}\pi$ and in (b) $\theta=\frac{1}{5}\pi $. {{These density matrices have similar discord, as can be seen from Fig.~\ref{Discord},  yet their visibility landscapes look completely different: an almost $\pi $-jump in zero-visibility lines over a small interval of $\beta $ followed by an almost $\beta $-independent zero-visibility lines in (a) is equivalent to  zero-visibility lines with a small non-monotonicity over a large interval in (b); both signify weak sensitivity with respect to changes in passive system for the states with a relatively small discord.}} }
\end{figure*}

 {Here we introduce a new discord quantifier and show how it works on examples of protocol  implementation for known states. Let us   start with specifying simple \emph{real}  input states for both subsystems, i.e.\ choosing $\phi _\nu=0$ in Eq.~\eqref{X-nu} so that each of these states can be written as $\ket{A_\nu}\equiv \ket{\theta_\nu}=\cos \tfrac{1}{2}\theta_{ \nu} \ket{\uparrow}+ \sin\tfrac{1}{2} \theta_{ \nu}\ket{\downarrow}  $, i.e.\  parameterized only via a single parameter, $\theta_\nu^A\equiv \theta_\nu$. Similarly, each $\ket{B_\nu} $ is parameterized only via a single parameter $\theta_\nu^B$. Next choose $\phi _{A,B}=0$ in the evolution matrices ${\mathsf{S}}^{A,B} $, Eq.~\eqref{SB}. In this case, lines of constant visibility  for a given in-state   are functions of $\phi _d$ and the two parameters, $\alpha $ and $\beta $, describing quantum evolution through the subsystems $A$ and $B$. As the auxiliary state $\ket{A} $ in Eq.~\eqref{ABa}, and hence diagonalizing matrix ${\mathsf{S}}_0^A$ in Eq.~\eqref{S0A},  are also real, the zero-visibility lines correspond to $\phi _d=0$.}

{For real in-states, there could be no more than two linearly independent sets for each subsystem.  Choosing these sets to be `symmetric',     $\theta_{ \nu}^B={\theta}_\nu$, leads to the parameterization $\rho^{AB}=\sum_{\nu=1}^2 w_\nu\ketbra{\theta_{ \nu}  {\theta}_\nu;0 }$, i.e.\ each in-state is defined by only three parameters, $\theta_1,\, \theta_2 $ and $w_1$ (with $w_2=1-w_1$).  Here and elsewhere, we use the following notations for partial states of the composite system \footnote{The states in Figs.~\ref{fig2} and \ref{fig3} are real, but in Appendix B we consider those with non-zero phases in active subsystem A; although it is straightforward to include a phase difference in  states in passive subsystem $B$, for simplicity we assumed it to be zero, thus omitting $\phi _\nu^{B} $ from the notation \eqref{ABnu}}:}
\begin{align}\label{ABnu}
\rho_\nu^{B}\otimes\rho_\nu^B =\ketbra{A_\nu B_\nu} \,,\quad\ket{A_\nu B_\nu}\equiv \ket{\theta_\nu^A,\theta_\nu^B; \phi _\nu^A} .
\end{align}

In  Fig.~\ref{fig2}, we present the visibility landscape for two particular choices of the  parameters: $\theta_1=0$, $\theta_2=\frac{\pi }{2}$ for state ({a}) which has the maximal  $A$-discord,  and   $\theta_1=\frac{\pi }{2}$, $\theta_2=-\frac{\pi }{2}$ for state ({b}) which has zero $A$-discord, with $\phi _\nu=0$ and  $w_1=\frac{1}{2}$ in both these cases.

{The dependence  $\alpha _0({\beta })$ corresponding to  the zero-visibility lines in the landscapes of Fig.~\ref{fig2}  reveals a striking difference between the non-discorded and discorded states: the latter shows a strong dependence on $\beta $ while the former is $\beta $-independent; this certainly  works not only for the chosen but for generic mixed states.}

\begin{figure}
  \includegraphics[width= .9\columnwidth]{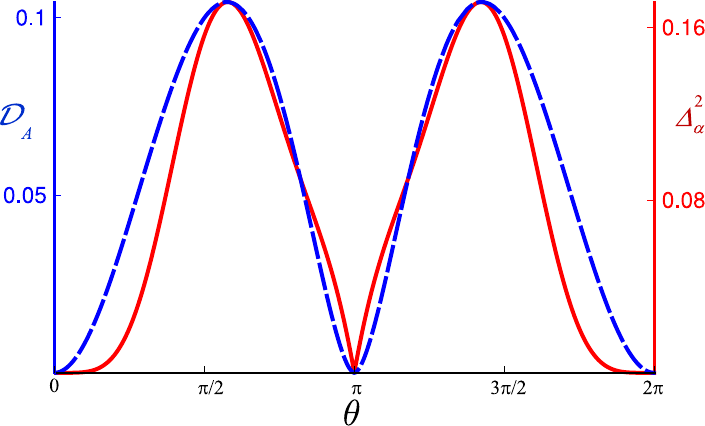}\\
  \caption{The standard definition of discord, $\mathcal{D}_A$, (dashed, blue online) vs the  alternative quantifier of Eq.~\eqref{W}, $\Delta^2_\alpha $, (solid, red onine)   for the in-state with the density matrix $\rho^{AB}_\theta=\frac{1}{2}\left[\,\ket{\uparrow\uparrow}\bra{\uparrow\uparrow}+ \ket{\theta\theta}\bra{\theta\theta} \,   \right]$. }\label{Discord}
\end{figure}

\begin{figure*}
\qquad
\includegraphics[height=.44\textwidth]{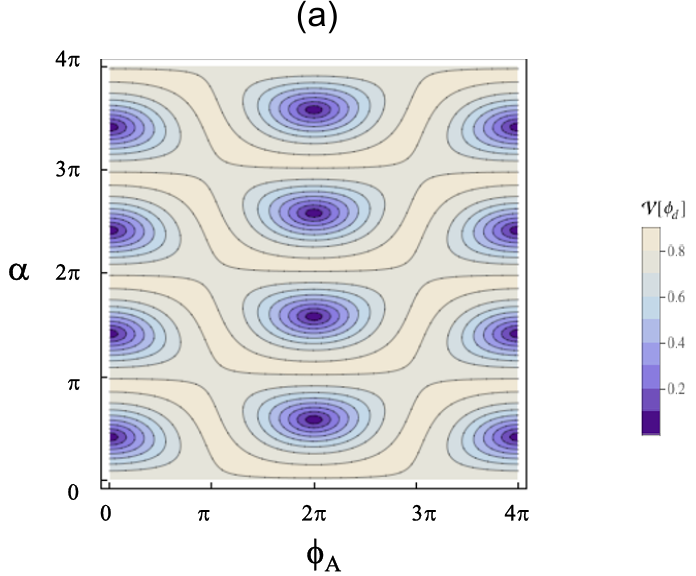}\quad
\includegraphics[height=.43\textwidth]{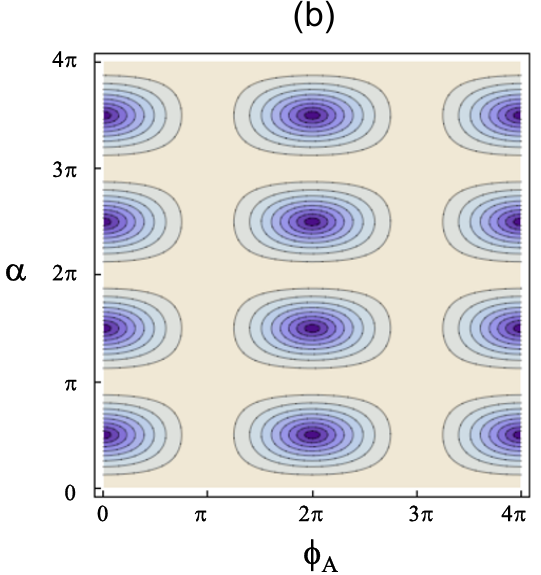}
\vspace*{-6pt}

\caption{ \label{fig5} The raw visibility landscape for the two in-states used in Fig.~\ref{fig2} as a function of parameters $\alpha $ and $\phi _A$ controlling, respectively, the transparency of BS$_1^A$ and the phase difference in   MZI$^A$ (see Fig.~\ref{MZI}). Here we keep fixed the  values of corresponding parameters in MZI$^B$ ($\beta =\pi/3$ and $\phi _B=0$). Here zero-visibility points correspond to $\phi _0=0\operatorname{mod}(2\pi) $ as expected. }
\end{figure*}
  The eye-catching signature of discord in Fig.~\ref{fig2}(a) is a high non-monotonicity of the zero-visibility lines, $\alpha _0({\beta })$.  However, a  $\pi $-periodic in $\alpha $ pattern of the zero-visibility lines implies that  vertical $\pi $-jumps in zero visibility curves happen for non-discorded states. Hence,   nearly $\pi $-jumps in a zero-visibility curves over a small interval of $\beta $, Fig.~\ref{fig3}(a), signifies weak sensitivity with respect to changes in the passive subsystem similar to that in curves with a small non-monotonicity over a large interval, Fig.~\ref{fig3}(b).  To treat both cases on equal footing,  we  employ the standard deviation of $f_\alpha({\beta})\equiv \cos^2[{\alpha_0({\beta})}]$ from its average over the period as a {quantifier} of such a   sensitivity, which plays the role of a \emph{discord quantifier}:
\begin{align}\label{W}
  \Delta^2_\alpha&=\int\limits_{0}^{2\pi}\frac{\mathrm{d}\beta}{2\pi} \left[f_\alpha({\beta}) -\overline{f}_{\!\alpha}\right]^2\!, &
   \overline{f}_{\!\alpha}&=\int\limits_{0}^{2\pi} \frac{\mathrm{d}\beta}{2\pi}f_\alpha({\beta}).
\end{align}

 This quantifier gives similar results for the two sets of symmetric in-states in Fig.~\ref{fig3}. Both have the density matrix  \mbox{$\rho^{AB}_\theta=\frac{1}{2}\left[\ket{\uparrow\uparrow}\bra{\uparrow\uparrow}+ \ket{\theta\theta}\bra{\theta\theta}    \right] $} with different $\theta$. For $\theta\!=\!0 $, $ \rho^{AB} =\ket{\uparrow\uparrow}\bra{\uparrow\uparrow}$ is a pure state  with no discord, and likewise discord is  absent for  $\theta=\pi $ when $\rho^{AB}_\theta\to\frac{1}{2}\left[\,\ket{\uparrow\uparrow}\bra{\uparrow\uparrow}+ \ket{\downarrow\downarrow}\bra{\downarrow\downarrow} \,  \right]$. Thus, discord is small for $\rho_\theta^{AB} $ with $\theta$ approaching either $0$ or $\pi $, cf.\ Fig.~\ref{Discord}.

This suggested quantifier is convenient and, although it is  by no means unique, it works remarkably well:  its similarity to  quantum discord in its original definition  is    quite  appealing, as illustrated for $\rho_\theta^{AB} $ of the above example  in Fig~\ref{Discord}.    It is straightforward to prove that this measure is reliable:   it  vanishes for any non-discorded state and does not change with a unitary transformation on passive subsystem $B$.

{In Appendix B we give further examples of discorded and non-discorded states, including that with non-zero phases and that where the density matrices in Eq.~\eqref{sepstate} are spanned by more than two states.  We also describe there a useful generalization of the discord quantifier for in-states with non-zero phases.}
\section{Employing the Protocol  for Unknown States\label{section5}}
Experimentally, any in-state, Eq.~\eqref{sepstate}, is \emph{repeatedly} generated  in the scheme given in Fig.~\ref{MZI} by  random {simultaneous} changes of transparencies of beam-splitters BS$^B_0$ and BS$^A_0$ with fixed probabilities $w_\nu$.  A set of raw data for the generated in-state should be obtained by varying the phase difference, $\phi _d$, in the detecting MZI$^{d} $ and  measuring the appropriate particle cross-correlation function, Eq.~\eqref{K}.  From this data set, one extracts the visibility ${\mathcal{V}}$ defined in step 6 of the protocol, Section \eqref{section2}. Fixing the phase difference  $\phi _B$ in the passive subsystem $B$ makes ${\mathcal{V}}$  a function of three  parameters that experimentally control the in-state evolution through the system:  $\alpha $ and $\phi _A$, characterizing the scattering matrix ${\mathsf{S}}^A$, Eq.~\eqref{SB}, and $\beta $ characterizing  ${\mathsf{S}}_B$.

Fixing also $\beta $, one represents the data as  lines of constant visibility in the $\alpha -\phi _A$ plane, thus producing the visibility landscape.   From this one finds $\phi _{A0}$ and $\alpha _{0}$ that correspond  to zero visibility for this value of $\beta $.  Repeating this for different values of $\beta $, one derives the parametric representation of the zero visibility lines as $\alpha _0({\beta })$ and $\phi _{A0  }({\beta }) $. {This step was not required in the example of Fig.~\ref{fig2}, as in  such a case of  real in-states one expects $\phi _A=0$. Indeed, from the visibility landscape (where visibility lines are drawn as functions of $\alpha $ and $\phi _{A}$) for the in-states used in this example, Fig.~\ref{fig5}, one clearly sees that zero-visibility points correspond to $\phi _{A0}=0\operatorname{mod}(2\pi) $ as expected. Hence, $\alpha _0({\beta })$ dependence alone is sufficient for quantifying discord for such states, \eqref{W} and Fig.~\ref{Discord}.}

{For a generic (unknown) in-state characterized by arbitrary phases, one should first build a visibility landscape in the $\phi _A$-$\alpha$ plane in order to determine the values of the phase $\phi _A$ corresponding to the zero visibility.  Fixing these values, one then builds the corresponding visibility landscape on the $ \alpha-\beta $ plane and uses this   for   the discord detection and its full characterization via the correlation discord quantifier. In Appendix B, we illustrate how this works using known in-states with a non-zero phase.}

\section{Conclusion}
 We   have proposed  a new characterization of quantum discord based on measuring cross-correlations in non-entangled bipartite systems and thus linear in density matrix $\rho$,  in contrast to other quantifiers, notably geometric discord, that require full or partial quantum tomography for reconstruction of $\rho$. The linearity of the proposed quantifier opens a path to extending experimental research of discord into electronic condensed matter systems. We have considered in detail one possible implementation via devices built of Mach -- Zehnder interferometers in quantum Hall systems, where our quantifier
 is quite robust against external noise and fluctuations: as long as the Aharonov--Bohm oscillations are resolvable \cite{Heiblum:03}, the appropriate interference pattern may serve as a pictorial discord witness, as illustrated above in Figs.~\ref{fig2} and \ref{fig3}. Finally, our discord quantifier  is qualitatively consistent, and quantitatively very close  to the original measure.

 The relative simplicity of this protocol, and the fact that it is based on presently existing measurement technologies and available setups (electronic Mach-Zehnder interferometers) is bound to stimulate experiments in this direction. While the present
analysis addresses discord of bipartite systems, an intriguing generalization of our protocol to multiply-partite systems is possible by introducing a number of coupled interferometers
 Extension of our protocol to anyon-based states (employing anyonic interferometers) or other topological states may open the horizon to topology-based study of discord.

\begin{acknowledgements}This work was supported by the Leverhulme Trust Grants RPG-2016-044 (IVY), VP1-2015-005 (IVY, YG), the Italia-Israel
project QUANTRA (YG), and the DFG within the network CRC TR 183, C01
(YG). The authors (IVL, IVY and YG) are grateful for hospitality extended to them at the final stage of this work at the Center for Theoretical Physics of Complex Systems, Daejeon, South Korea.
\end{acknowledgements}

\begin{appendix}
\begin{figure*}
\includegraphics[height=.45\textwidth]{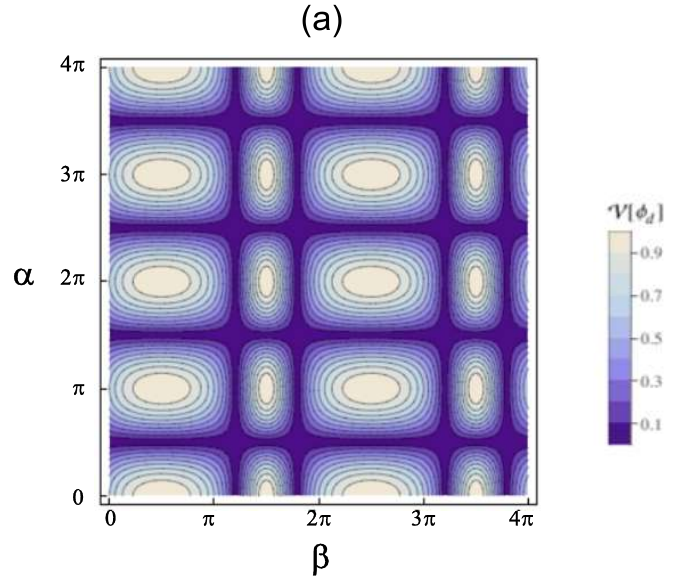}\quad
\includegraphics[height=.45\textwidth]{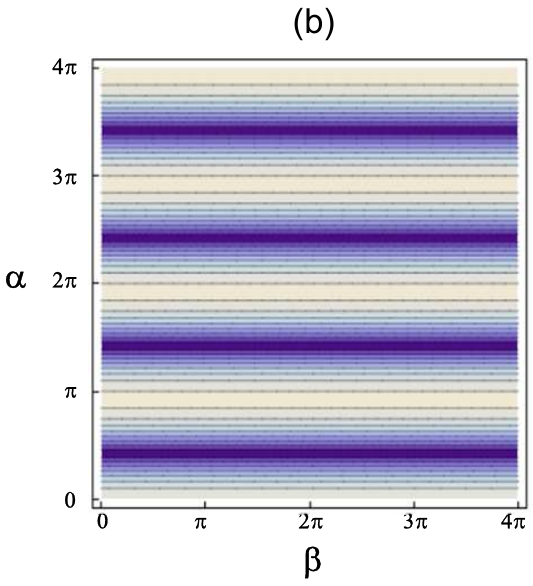}
\vspace*{-6pt}

\caption{Visibility as a function of $\alpha$ and $\beta$ for non-discorded states with $\phi _A=\phi _{1,2}$. (a): $\rho^{AB}=1/5\ket{++}\bra{++}+4/5\ket{--}\bra{--}$ (b): $\rho^{AB}=1/2\ket{\uparrow\uparrow}\bra{\uparrow\uparrow}+1/2\ket{+\downarrow}\bra{+\downarrow}$. The first plot (a) displays the 'grid-like' visibility characteristic of a density matrix which is correlated between $A$ and $B$ subsystems, but non-discorded, with only the classical correlations between subsystems. The `barcode' graph (b) is a result of a density matrix which is completely uncorrelated between $A$ and $B$ subsystems. } \label{fig6}
\end{figure*}
\begin{figure*}
\includegraphics[height=.45\textwidth]{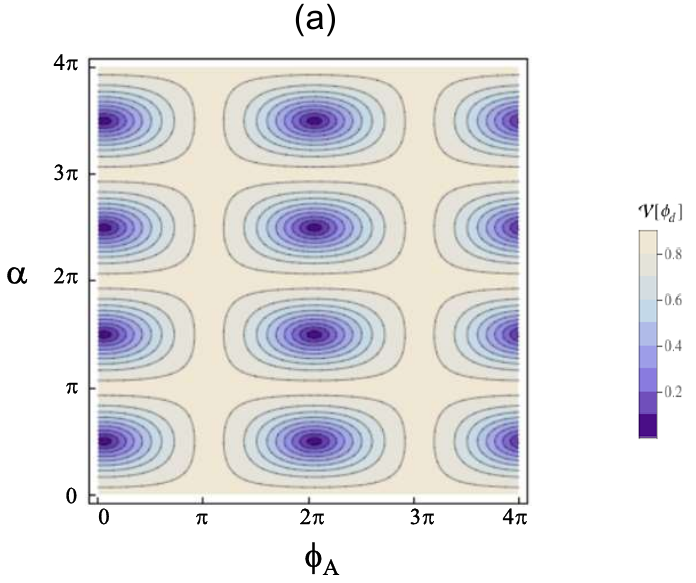}\quad
\includegraphics[height=.45\textwidth]{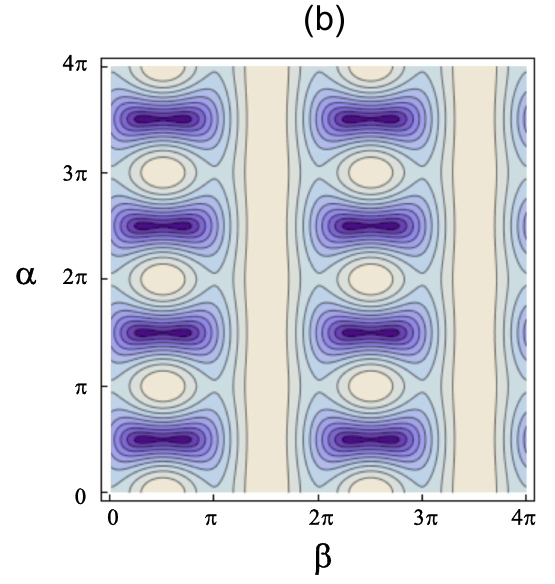}
\vspace*{-6pt}

\caption{  The visibility landscapes for the density matrix given by Eq. (\ref{phase} with   $\phi_2=\pi/2$: (a) Visibility as a function of $\alpha,\phi_A$ with $\beta=2\pi/3$; (b)  Visibility   as a function of $\alpha,\beta$ with fixed $\phi_A=\arctan\big(\frac{2-\sqrt{3}}{3}\big)$  that corresponds to minimal visibility spots in the landscape plot (a).}\label{fig7}
\end{figure*}

\section{Quantum Discord}\label{App1}
Quantum Discord \cite{Zurek:01,Vedral:01} exemplifies the difference between classical and quantum correlations of two subsystems, {$A$ and $B$}, as quantified by mutual information. The latter, which is a classical measure of correlations between  {$A$ and $B$}, is defined   as ${I}(A{:}B)\equiv H(A)+H(B)-H(AB)$, where    the Shannon entropy $H(A)\equiv-\sum_a p_{a}\log p_{a}$ with $a$ being the possible values that a classical variable $A$ can take with the probability $p_a$, while the joint entropy $H(AB) $ is that of the entire system $A\bigcup B$. An alternative way of writing a classically equivalent expression to $I(A{:}B)$ is $J(A{:}B)\equiv H(A)-H(A|B)$, with $H(A|B)\equiv H(AB) -H(B)$ being the conditional entropy which is the uncertainty remaining about $A$ given a knowledge of $B$'s distribution.

The quantum analogues to these expressions can be obtained  by replacing the Shannon entropies for the probability distributions with the corresponding von Neumann entropies for QM density matrices, $S(\rho )=-\Tr\{\rho \log\rho \}$. The quantum analogue of $I(A{:}B)$ is then straightforward to define,
\begin{align}
\mathcal{I}(\rho^{AB})&\equiv S(\rho^A)+S(\rho^B)-S(\rho^{AB}), \label{I(rhoAB)}
\end{align}
where $\rho^A,$ $\rho^B$ are the reduced density matrices on either subsystem. However, the straightforward analogue to the classical conditional entropy is not that useful:  if one defines $S({B|A})=S({AB})-S({A})$, this quantity could be negative, e.g., in the case when   subsystems $A$ and $B$ are in a pure state. Instead, the quantum conditional entropy $S({A|B})$ is defined as the average von Neumann entropy of states of $A$ after a measurement is made on $B$.

The result of a measurement depends on the basis   picked for the measurement projectors. Post-measurement density matrix becomes
\begin{align}
\widetilde\rho^{AB}=\sum_{\mu}\,p_{\mu}^A\,\Pi_{\mu}^A\otimes\rho_{B|\Pi^A_{\mu}}
\end{align}
where $\rho_{B|\Pi^A_{\mu}}$ is the density matrix conditional on some measurement on $A$ defined as follows,
\begin{align}
\begin{aligned}
\rho_{B|\Pi^A_{\mu}}&\equiv\frac{1}{p_{\mu}^A}\,
\Tr_{A}\left(\Pi^A_{\mu}\otimes{\mathbb 1}^B\right)\rho^{AB}\left(\Pi^A_{\mu}
\otimes{\mathbb 1}^B
\right)\,
\\ p_{\mu}^A&=\Tr \left(\Pi^A_{\mu}\otimes{\mathbb 1}^B\right) \rho^{AB}
\end{aligned}
\label{RhoCon}
\end{align}
Using this conditional state, Eq.~(\ref{RhoCon}), one may extract the entropy $S(\rho_{B|\Pi^A_{\mu}})$ which gives us the amount uncertainty of the state of $B$ given this projection of $A$ into a measurement basis. Then   the conditional entropy after a complete set of measurements   $\{\Pi^A_{\mu}\}$ becomes
\begin{align}
S(B|\{\Pi^A_{\mu}\})\equiv\sum_{\mu} p^A_{\mu} S(\rho_{B|\Pi^A_{\mu}})\,.
\end{align}
Now a generalization of $J(A{:}B)$ can be constructed,
\begin{align}
\mathcal{J}_A(\rho^{AB})\equiv S(\rho^B)-\max S(B|\{\Pi^A_{\mu}\}),
\end{align}
where one final ingredient has also been added in order to remove the dependence on the measurement basis:   maximizing over all complete measurement bases, essentially equivalent to picking the \emph{best} measurement basis (that is the one where the ignorance about subsystem $A$ is reduced the most).

Having defined two quantities which would be classically equivalent, the difference between the two could be thought of as a measure of quantumness. It is the quantity which is termed the \emph{quantum discord}:
\begin{align}\label{QDdef}
\begin{aligned}
\mathcal{D}_A(\rho^{AB})&\equiv\min_{\{\Pi^B_{\mu}\}}
\left[\mathcal{I}(\rho^{AB})-\mathcal{J}_B(\rho^{AB})\right]\\&=
\min_{\{\Pi^B_{\mu}\}}S(A|\{\Pi^B_{\mu}\})-\left[S(\rho^{AB})-S(\rho^B)\right].
\end{aligned}
\end{align}

Note that since $\mathcal{J}$ is not symmetric about which subsystem the measurement is performed on, neither is discord and in general $\mathcal{D}_B(\rho^{AB})\neq \mathcal{D}_A(\rho^{AB})$.

\begin{figure*}
\includegraphics[width=.38\textwidth]{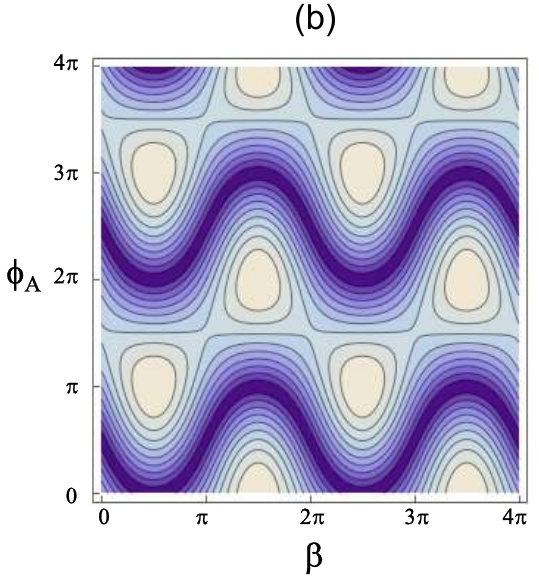}\qquad\quad
 {\includegraphics[width=.52\textwidth]{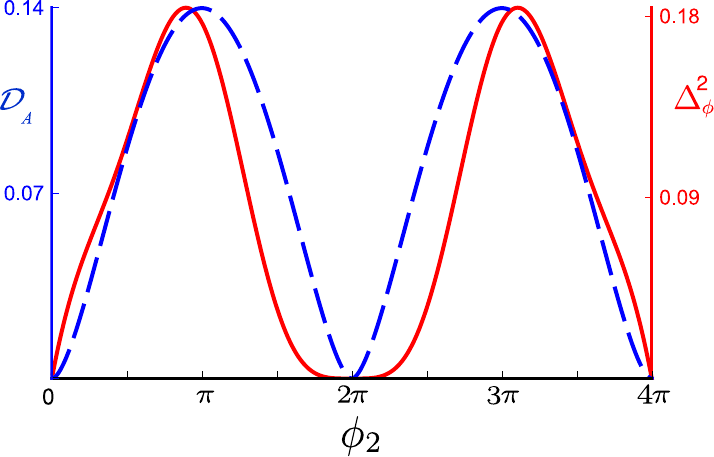}}
\vspace*{-6pt}

\caption{(a) Visibility landscape for the density matrix given by Eq.~(\ref{phase} with  $\phi_2=\pi/2$, where visibility lines are dependent  on $\phi_A$ and $\beta $ with fixed diagonalising parameter  $ \alpha  =\alpha _0=\pi/2$. (b) Discord (dashed, blue online) and an alternative quantifier, $\Delta^2_\phi$, (red, solid online) for the state (\ref{phase} for a range of $\phi_2$.}\label{fig8}
\end{figure*}
\begin{figure*}
\begin{minipage}{0.45\textwidth}\includegraphics[width=\textwidth]{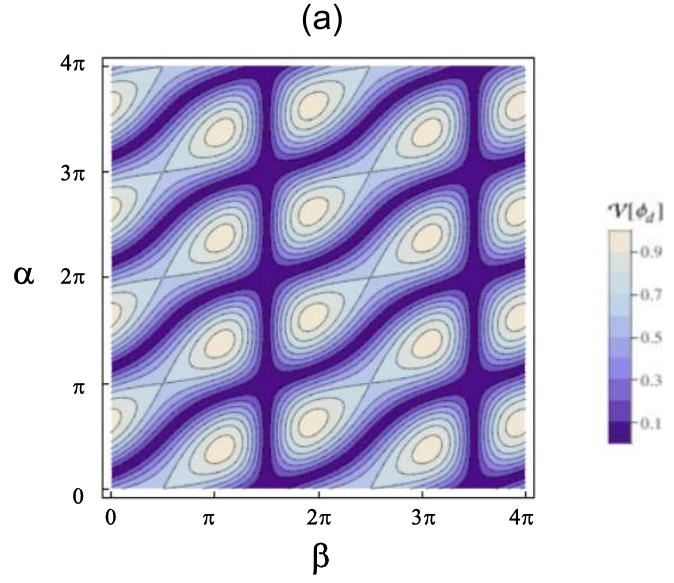}\end{minipage}
\begin{minipage}{0.45\textwidth}\includegraphics[width=\textwidth]{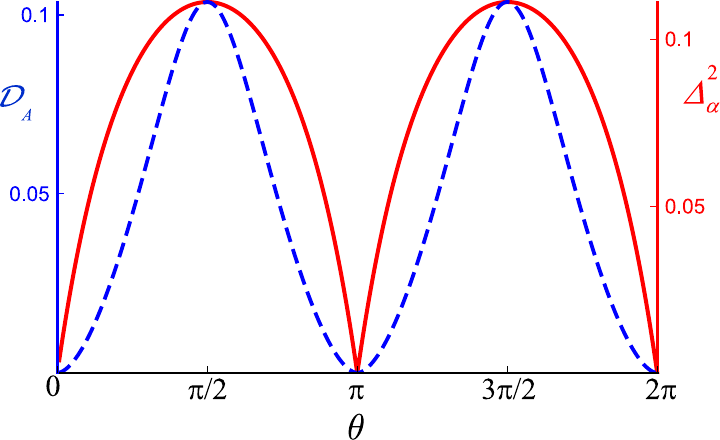}\end{minipage}
\vspace*{-6pt}
\caption{(a) Visibility landscape for the density matrix given by Eq.~(\ref{3}) with $\theta=\pi/2$, with a characteristic signature of discord  in the curviness of the zero-visibility lines; (b) the discord quantifier  (red, solid online) of Eq.~\eqref{W} vs the standard discord (dashed, blue online) for the density matrix given by Eq.~(\ref{3}) for a range of $\theta$.} \label{fig:3states}
\end{figure*}

\section{Further examples of discord characterization via visibility landscapes}
The `grid-like' non-discorded state of Fig.~\ref{fig2}(b) corresponds to maximal possible \emph{classical} correlations between the subsystems. In such a case  no information about the correlations between subsystems $A$ and $B$ is lost when one makes the correct choice of measurement on subsystem $B$. {Actually, any classically-correlated states with no discord would look grid-like. We give another example of an non-discorded in-state of Eq.~\eqref{sepstate} with $n=2$,  choosing there $w_1=1/5$ and  define $\rho_\nu^A$ via states $\ket{A_{1,2}} $, Eq.~\eqref{X-nu} where we put $\theta_1=-\theta_2=\pi /4$ and $\phi _1=\phi _2$. In this case, although the states $\ket{A_\nu} $ are complex, a relative phase between them is zero. If such a state were unknown, one would find from the $\alpha $-$\phi _A $ plot that the zero-visibility spots correspond to $\phi _A=\phi _{1,2}$. Fixing this value of $\phi _A$ results in the grid-like plot on the $\alpha $-$\beta $ plane, Fig.~\ref{fig6}(a), clearly shown the absence of discord. When not only discord but classical correlations  between $A$ and $B$ subsystems are also absent, the visibility lines become `barcode-like', i.e.\ only horizontal, Fig.~\ref{fig6}(b).}

{If we choose an in-state with the same characteristics as the non-discorded one in Fig.~\eqref{fig2}(b) but different phases in subsystem $A$,   i.e.\
\begin{align}\label{phase}
    \rho^{AB}&=\tfrac{1}{2} \left[\ketbra{\bm {++};0}+\tfrac{1}{2} \ketbra{\bm {--};\phi _2}\right]
\end{align}
then such a state is $A$-discorded provided that \mbox{$\phi _2\ne0(\operatorname{mod}\pi )$}, with  discord reaching the maximum at $\phi_2=\pi/2$. However, the quantifier of Eq.~\eqref{W} is not sufficient for its full description since  the values of  $\phi _A$ where visibility drops to zero are now $\beta $-dependent themselves. To illustrate this, we build the visibility landscape in $\alpha $-$\phi _A$ axes for the maximally discorded state for different values of $\beta $, as illustrated in Fig.~\ref{fig7}(a) for $\beta =2\pi /3$. Extracting   $\phi _{A0}({\beta })$ corresponding to zero visibility points (which is given analytically for this particular known state by $\phi_{A0}=\arctan(2/3-1/\sqrt{3}) $ but can, in general, be found from the plot), we build $\alpha $-$\beta $ visibility landscape shown in   Fig.~\ref{fig7}(b). }%\newpage

{It becomes immediately obvious that the quantifier of Eq.~\eqref{W} is not at all convenient in this case. Although the state \eqref{phase} is always $A$-discorded for $\phi _2=\pi /2$, there are regions  where the state can not be diagonalised and  zero-visibility lines, on which the quantifier \eqref{W} is based, are absent.  The reason is that now there are two diagonalising parameters, $\alpha _0$ and $\phi _{A0}$, and it is their joint dependence on the parameters of passive subsystem $B$ that fully reveals and characterizes discord. For this particular example, it is $\phi_{A0}({\beta })$-dependence alone that describes discord practically in full, as illustrated in Fig.~\ref{fig8}. There we have introduced, similar to Eq.~\eqref{W}, the standard deviation of $f_\phi ({\beta})\equiv \cos^2[{\phi _{A0}({\beta})}]$ from its average over the period as a {quantifier} of the sensitivity of diagonalising parameters to changes in passive subsystem $B$, with the only difference that it is $\phi _{A0}$ rather than $\alpha _0$ which is now the variable diagonalising parameter:}
\begin{align}\label{W2}
  \Delta^2_\phi&=\int\limits_{0}^{2\pi}\frac{\mathrm{d}\beta}{2\pi} \left[f_\phi({\beta}) -\overline{f}_{\!\phi}\right]^2\!, &
   \overline{f}_{\!\phi}&=\int\limits_{0}^{2\pi} \frac{\mathrm{d}\beta}{2\pi}f_\phi({\beta}).
\end{align}

{In general, it is the sum $\Delta_\alpha ^2+\Delta_\phi ^2$ that fully characterizes $A$-discord of a complex in-state. In order to experimentally obtain $\Delta^2_\alpha+\Delta_{\phi}^2$ for an unknown state, one builds the full zero-visibility lines in three-dimensional parameter space $(\alpha,\,\beta,\,\phi_A)$. The discord quantifier is extracted from  this line by calculating $\Delta^2_\alpha+\Delta_{\phi}^2$, which is zero only if  discord is absent. The separate measures $\Delta^2_\alpha$  and $\Delta^2_\phi$ can be obtained by the projection of the line onto the $\alpha-\beta$ and $\phi_A-\beta$ planes respectively.}

For a final illustration, we present an example of in-state with $n=3$ in Eq.~\eqref{sepstate}. We choose a real in-state
\begin{align}\label{3}
    \rho^{AB}=\tfrac{1}{3}\ketbra{\uparrow\uparrow}+\tfrac{1}{3}
    \ketbra{\downarrow\downarrow}+\tfrac{1}{3}\ketbra{\theta\theta} %
\end{align}
As it is real, the diagonalising value of $\phi _A$ is zero, so that the visibility landscapes as functions of $\alpha $ and $\beta $ allow one to extract the discord quantifier of Eq.~\eqref{W} strikingly similar to the standard definition of discord, see Fig.~\ref{fig8}.

\end{appendix}

%\bibliography{my}\end{document} 

%

\end{document}